\documentclass[twocolumn,prd,preprintnumbers,amsmath,amssymb]{revtex4}
\usepackage{graphicx}
\usepackage{times}
\usepackage{epsfig}
\def\pa{\partial}

\def\dF{^{\star\!}F}

\def\cald{{\cal D}}

\def\call{{\cal L}}
\def\calm{{\cal M}}

\def\calu{{\cal U}}

\def\beq{\begin{equation}}
\def\ee{\end{equation}}
\def\eeq{\end{equation}}
\def\pa{\partial}

\def\bfig{\begin{figure}}
\def\efig{\end{figure}}
\def\bea{\begin{eqnarray}}
\def\beann{\begin{eqnarray*}}
\def\eea{\end{eqnarray}}
\def\eeann{\end{eqnarray*}}

\def\nn{\nonumber}

\def\3p0{$^{3}P_{0}$}

\begin{document}


\title{CP Violation  in  Dual Dark Matter
}

\author{ C. A. Z. Vasconcellos, D. Hadjimichef}
\affiliation{
Instituto de F\'{\i}sica, Universidade Federal do Rio
  Grande do  Sul\\
 Av. Bento Gon\c{c}alves, 9500\\ Porto Alegre, Rio Grande do Sul,
CEP 91501-970, Brazil
}

\begin{abstract}
 In this paper   we  study the consequences of extending
dual symmetry (DuSy) to include a generic $C_\mu$ vector field as
a dual partner of the photon $A_\mu$. A new combined field, the
complex  $Z^\prime_\mu$, is obtained from  $C_\mu$ and $A_\mu$.
The promotion of dual symmetry to a local symmetry for
$Z^\prime_\mu$   implies in the inclusion of an extra complex
vector field $W_\mu$ with a complex gauge transformation. A  dual
dark matter (DM) Lagrangian $\call_{\rm DDM}$ is obtained from the
general DuSy invariant Lagrangian $\call_{\rm DuSy}$. Our
tentative conjecture is to interpret $W_\mu$ as the actual weak
interaction charged gauge boson $W^{\pm}_\mu$, which leads us to
speculate about a possible extra CP violation scenarios for future
calculations.
\end{abstract} 
\maketitle

\section{Introduction}

The enigma of dark matter (DM) remains unsolved, a mystery which
has been puzzling scientists for more than 80 years.  Still one of
the most evasive and fascinating mysteries in physics,  the
problem of the DM in the Universe stirs the imagination of  most
astronomers, cosmologists and particle physicists, that are
convinced that at least 90\% of the mass of the Universe is due to
some non-luminous form of matter.

In 1933, the astronomer Fritz Zwicky (Zwicky 1933) has observed
the radial velocities of eight galaxies in the Coma cluster,
providing the first evidence for DM. He has found an unexpected
large velocity dispersion of these galaxies which would indicate
that the visible mass of (luminous) matter in the Coma cluster was
much smaller than its total mass. From these observations, he has
found that the mean density of the Coma cluster would have to be
$400$ times greater than the one derived from luminous matter.
Zwicky then concluded that DM (non luminous matter) would be
present and that the large velocity dispersion in Coma (and other
groups of galaxies) represents an unsolved problem.

In 1970, Rubin and Ford (Rubin \& Ford 1970) obtained the
strongest evidence up to that time for the existence of DM which
became the most plausible explanation for the anomalous rotation
curves in spiral galaxies.

Cosmology and astrophysics have provided historically, many
convincing evidence for the existence of DM. The primordial
observations which revealed the lack of mass to explain the
internal dynamics of clusters of galaxies and the rotation of
galaxies, which date back as we have seen the years 30 and 70,
have been followed, more modernly, by observations that have
provided substantial and consistent evidence for the existence of
mass effects in regions of the universe where no luminous mass is
observed.


 Although the existence of DM is widely
 accepted, several other explanations of these discrepancies have also been proposed.
Still, besides these signs of gravitational effects, there are
other types of searches for dark matter, as the identification of
explicit manifestation of DM particles and indirect searches
involving annihilation or decay of DM particles in the flow of
cosmic rays (Cirelli 2013; Bertone 2010).

Past all these years after  the original suggestion  related to
its existence, we still do not know the composition of dark matter
or its nature. Global fits to gravitational signal effects of DM
at the cosmological and extra galactic  scale for a large range of
data sets (cosmic background radiation, large-scale structure of
the universe, type Ia supernovae) determine with great accuracy
the amount of cold DM in the overall content of matter energy in
the universe at $\Omega_{DM} h^2 = 0.1123 \pm 0.0035$ (Cirelli
2013), where  $\Omega_{DM}\,h^2$ represents the density of cold DM
matter and  $h=H_0 /100$ km s$^{-1}$ Mpc$^{-1}$ is the Hubble
parameter. Moreover, the observations after nine years release
have placed stringent constraints to the abundance of baryons (B)
and dark energy (DE) (Bennett et al. 2003; Spergel et al. 2003;
 Komatsu et al. 2009):
$ \Omega_B\,h^2 = 0.233 \pm 0.023 \, \, ; \, \Omega_{DE}\,h^2 =
0.721 \pm 0.025 $ where $\Omega_B h^2$ is the physical baryon
density and $\Omega_{DE} h^2$ is the dark energy density.

Modernly, there are many plausible candidates for dark matter, as
{\it weakly-interacting massive particles} (WIMPs), SM neutrinos
(Weinheimer 2003), sterile neutrinos (Dodelson \& Widrow 1994),
axions (Rosenberg \& Bibber 2000), supersymmetric candidates
(neutralinos, sneutrinos, gravitinos, axinos (Falk, Olive,  \&
Srednicki 1994; Feng, Rajaraman \& Takayama 2003; Goto \&
Yamaguchi 1992),
 light scalar DM (Lee \&  Weinberg 1977), DM from little Higgs models (Cheng \& Low 2003),
Kaluza-Klein particles (Agashe \& Servant 2004), superheavy DM
(Griest  \& Kamionkowski 1990). An excellent review in theoretical
and experimental aspects of DM can be found in Bertone, Hooper \&
Silk (2005). In general they are present in theories of weak-scale
physics beyond the Standard Model (SM) and give rise to
appropriate relic abundance. Calculations have shown that stable
WIMPs can remain from the earliest moments of the Universe in
sufficient number to account for a significant fraction of relic
DM density. This raises the hope of detecting relic WIMPs directly
by observing their elastic scattering on targets. In the DM zoo
many different types of particles have been introduced and their
properties theoretically studied.

 Many types of models that explore the
physics beyond SM, share in common the presence of new
$U(1)$ vector bosons. These new bosons are introduced
basically in two ways: (i) minimal coupling; (ii) Stueckelberg mechanism. A new vector gauge boson would be
massless if a new $U(1)$ symmetry should remain unbroken. This would imply in a long range force if it were
to couple to ordinary matter, unless the coupling were
incredibly small. This case
would be allowed if the primary coupling were to a hidden
sector and connected only by higher-dimensional operators or alternatively by kinetic mixing with the photon.
In the case of kinetic mixing, this scenario would induce a
small fractional electric charge for hidden sector particles.
Usually this class of models the extra gauge boson is called the $Z^\prime$.
Now, the experimental discovery of a  $Z^\prime$ would be exciting,
but the implications would be much greater than
just the existence of a new vector boson. Breaking the
$U͑(1)^\prime$ symmetry would require, for example, an extended Higgs, with significant consequences
for collider physics and cosmology (Langacker 2009).

Following  a different approach than the  usual models with extra
$U(1)$ sectors in the SM, we study the consequences of the
introduction of  new complex vector bosons $Z^\prime$ and
$Z^{\prime\ast}$ subject to dual symmetry (DuSy) requirements. An
interesting conjecture is to apply this theory to CP violation. It
is well known that if CP violation in the lepton sector is
experimentally determined to be too small to account for
matter-antimatter asymmetry, some new Physics beyond the Standard
Model would be required to explain additional sources of CP
violation. Fortunately, it is generally the case that adding new
particles and/or interactions to the Standard Model introduces new
sources of CP violation. In what follows we use the Minkowski
space-time $r^{\alpha} =  (t, \mathbf{r})$ with signature ($-, +,
+, +$), and assume natural units $\epsilon_0 = \mu_0 = c = 1$.

\section{Dual Symmetry}

\subsection{A brief review}

Since seminal works of Dirac   on magnetic monopoles (Dirac 1931,
1948)  a great deal of  theoretical interest has appeared, despite
no confirmed experimental evidence of their existence up to the
present. Dirac's conclusions about the possibility of the
existence of magnetic monopoles were based on a logical
conclusion. In a paper published in 1931 (Dirac 1931), Dirac
showed that if a magnetic particle interacts with an electrically
charged particle, according to the laws of quantum mechanics, the
electric charge of the particle must necessarily be quantized.
From a complementary reciprocal reasoning, as we know that
electric charges are quantized Dirac concluded that magnetic
monopoles must be taken seriously.

In the article published in 1948 (Dirac 1948), Dirac sets up a
general theory of charged particles and magnetic poles in
interaction through the medium of the electromagnetic field.

 Using the four components coordinates $x_{\mu}$ with $\mu=0, 1,
3, 4$ to fix a point in space and time and the velocity of light
to be unit, the starting point of Dirac's formulation was the
ordinary electromagnetic field, $F^{\mu\nu}$ at any point: \bea
F_{\mu\nu}=\pa_\mu\,A_{\nu}-\pa_\nu\,A_{\mu} \,\,, \, \,\,  A = (V
, \mathbf{A}) \, ,  \label{F} \eea where $A_{\mu}$ represents the
electromagnetic $4$-potential.

In vacuum, Maxwell's equations
\begin{eqnarray}
\mathbf{\partial} \cdot \mathbf{E} = 0 ;  \, \mathbf{\partial} \!
\times \! \mathbf{E} = - \frac{\partial \mathbf{B}}{\partial t} ;
\, \, \mathbf{\partial} \! \cdot \! \mathbf{B} = 0 ; \,
\mathbf{\partial} \! \times \! \mathbf{B} = \frac{\partial
\mathbf{E}}{\partial t},
\end{eqnarray}
exhibit a high degree of symmetry: they are invariant under
conformal Lorentz transformation and under electromagnetic
duality, $(\mathbf{E},\mathbf{B}) \to (\mathbf{B},-\mathbf{E})$.
This discrete symmetry is a particular case of a continuous
symmetry $\mathbf{E} \to \mathbf{E} \cos \, \theta + \mathbf{B}
\sin \, \theta$ and $\mathbf{B} \to \mathbf{B} \cos \, \theta -
\mathbf{E} \sin \, \theta$, where $\theta$ is a pseudoscalar. A
consequence of this symmetry is that all fundamental properties of
free electromagnetic fields, such as energy, momentum, angular
momentum, among others, are symmetric with respect to this
transformation.

 Lorentz invariance can be made manifest by introducing
additionally to the electromagnetic field strength $F^{\mu\nu}$,
its dual $^*\!F^{\mu\nu}$.
 Since $F^{\mu\nu}$  is an antisymmetric
$6$-vector, $F^{\mu\nu} = - F^{\nu \mu}$, it can be shown that the
unique way to construct a dual Lorentz-invariant four-tensor
involving the electromagnetic field that is independent of the
original field strength tensor is
\begin{equation}
^*\!F^{\mu\nu} =  \frac{1}{2} \varepsilon^{\mu \nu \rho \sigma}
F_{\rho \sigma}, \label{dual}
\end{equation}
where $\varepsilon^{\mu \nu \rho \sigma}$ represents a tensor
equivalent to the Levi-Civita four-tensor. In Minkowski space,
${*}^2 = 1$. The duality transformation corresponds to $F^{\mu\nu}
\to ^*\!\!F^{\mu\nu}$, $^*\!F^{\mu\nu}  \to - F^{\mu\nu} $.

 The inner product $F^{\mu\nu}F_{\mu\nu}$
gives the Lorentz invariant
\begin{equation}
F^{\mu\nu}F_{\mu\nu} = 2 \left(B^2 - E^2  \right) \, ,
\end{equation}
where $E$ and $B$ represent respectively  the electric and
magnetic fields. The product of $F^{\mu\nu}$ with its dual
$^*\!F^{\mu\nu}$ allows to obtain the pseudoscalar Lorentz
invariant
\begin{equation}
^*\!F_{\gamma \delta}F^{\gamma \delta} = \frac{1}{2}
\varepsilon_{\alpha \beta \gamma \delta} F^{\alpha \beta}
F^{\gamma \delta} = - 4 \mathbf{E} \cdot \mathbf{B}\, .
\end{equation}

In the absence of field sources, it can be shown, that the
divergence of the dual field strength tensor is $\partial_{\nu} \,
^*\!F^{\mu \nu} = 0$. We can then express the physical content of
Maxwell's equations in a form which is elegant and manifestly
Lorentz-invariant, in the absence of field sources, as:
\begin{equation}
\partial_{\nu}F^{\mu \nu} = 0  \, \, ; \, \, \, \partial_{\nu} \, ^*\!F^{\mu \nu} =
0\, . \label{M}
\end{equation}
In Minkowski space, $\partial_{\nu} \, F^{\mu \nu} = 0$ implies
Eq. (\ref{F}). Similarly $\partial_{\nu}\,^*\! F^{\mu \nu} = 0$
implies \bea ^*\!F_{\mu\nu}=\pa_\mu\,\tilde{A}_{\nu}-\pa_\nu\,
\tilde{A}_{\mu} \,\,, \, \,\,  \tilde{A} = (\tilde{V} ,
\tilde{\mathbf{A}}) \, , \label{dualF} \eea where
$\tilde{A}_{\mu}$ represents the electromagnetic dual
$4$-potential. The duality transformation which relates $A_{\mu}$
and $\tilde{A}_{\mu}$ is nonlocal.
 In the presence of sources, duality invariance is
preserved provided we include in Eqs. (\ref{M}) electric
($j^{\mu}_e $) and magnetic sources ($j^{\mu}_m $) so that
\begin{equation}
\partial_{\nu}F^{\mu \nu} = j^{\mu}_e  \, \, ; \, \, \, \partial_{\nu} \, ^*\!F^{\mu \nu} =
j_m^{\mu}\, ,
\end{equation}
and the duality transformation is supplemented by a
cor\-res\-pon\-ding transformation of the sources
\begin{equation}
j^{\mu}_e  \to j^{\mu}_ m \, \, ; \, \,   j^{\mu}_m  \to -
j^{\mu}_e \, .
\end{equation}

The imposition of Lorentz invariance and invariance under duality
and parity transformations establishes severe restrictions on the
mathematical composition of the free Lagrangian density ${\cal
L}$: there are only three possibilities: $F^{\mu\nu} F_{\mu\nu}$,
$^{*}$$\!F^{\mu\nu}$$\,^{*}$$\!F_{\mu\nu}$ and
$F^{\mu\nu}$$\,^{*}$$\! F_{\mu\nu}$ at our disposal. The third
term is proportional to $\mathbf{E}. \mathbf{B}$ and is not
invariant under pa\-ri\-ty. The second term corresponds to the
dual sector of the original electromagnetic field and is not our
first option. We then write
\begin{equation}
 \call=-\frac{1}{4}\,F_{\mu\nu}\,F^{\mu\nu}\, . \label{max10}
\end{equation}
In this equation, the factor $-1/4$ is a matter of convention; any
multiplicative factor would have given the same equations of
motion.

In order to obtain a more complete Lagrangian density one might
try the naive choice
\begin{equation}
 \call=-\frac{1}{8} \left( F_{\mu\nu}\,F^{\mu\nu}
 + ^*\!\!F_{\mu\nu}\,^*\!F^{\mu\nu}\right) \, , \label{FF}
\end{equation}
where the two fields $F_{\mu\nu}$ and $ ^*\!F_{\mu\nu}$ are
considered independent. But since in reality they actually
describe the same electromagnetic field, ${\cal L} = 0$. The
second expression in Eq. (\ref{FF}) is actually a Bianchi
identity.

Another important property is that Maxell's equations are
invariant under the following dual global transformation \bea
F^{\,\mu\nu}&\to&\cos\theta\, F^{\mu\nu}+ \sin\theta\, \dF^{\mu\nu}\nn\\
\dF^{\,\mu\nu}&\to&-\sin\theta\, F^{\mu\nu} +\cos\theta\,
\dF^{\mu\nu}\, \label{max34} \eea where $\theta$ is the dual
angle. The inclusion  of matter, {\it i.e.} electric ($e$) and
magnetic ($m$) charges, implies that the zeros in (\ref{F}) are
replaced by electric $j_e^{\mu}$ and magnetic $j_m^{\mu}$ current
densities that obey  the following transformations \bea
j^e_{\nu}&\to&\cos\theta\, j^e_{\nu}+ \sin\theta\, j^m_{\nu}\nn\\
j^m_{\nu}&\to&-\sin\theta\,j^e_{\nu} +\cos\theta\, j^m_{\nu}\,,
\label{max35} \eea which guarantees that  dual symmetry is
preserved.

The Lorentz force written in a covariant notation \bea f^{\nu}=
j^{e}_{\mu}\,F^{\mu\nu}+j^{m}_{\mu} \,\dF^{\mu\nu} \label{imax36}
\eea is invariant under the combined transformations (\ref{max34})
and (\ref{max35}), so there is no fundamental restrictions on the
values of the dual angle $\theta$, which can be set to any
convenient value, as for instance,  \bea
\theta=\arctan\left(\frac{m}{e}\right). \label{imax37} \eea

Due to the duality transformation, one cannot decide whether a
particle has electric charge, magnetic charge  or both just
observing and comparing its behavior with the predictions of
Maxwell's equations. It is only a convention, not a requirement of
Maxwell's equations, that electrons have electric charge, but do
not have magnetic charge: a $\pi/2$ transformation of the dual
angle may originate exactly the opposite result. The fundamental
empirical fact is that all observed particles always have the same
ratio of magnetic charge and electric charge; transformations of
the dual angle can change the ratio to any arbitrary numerical
value but can not change the fact that all particles have the same
proportion of electric and magnetic charges. Thus, there must be a
transformation for which this reason equals to zero, so this
particular choice makes the magnetic charge equal to zero. This
particular choice is the conventional adopted de\-fi\-ni\-tion in
electromagnetism. For example, one can choose a dual
transformation that sets \bea j_m^{\mu}&\to&-\sin\theta\,j_e^{\mu}
+\cos\theta\, j_m^{\mu}=0\,, \label{imax38} \eea which results in
non-symmetrical Maxwell's equations
\begin{equation} \pa_{\nu}
F^{\mu\nu} = \frac{q}{e}j^{\mu} \nn \, ;  \, \,  \pa_{\nu}
\,\dF^{\mu\nu} = 0 \,, \label{imax39} \end{equation} with
$q^{2}=m^{2}+e^{2}$. In summary, applying a dual transformation on
symmetrical Maxwell equations (with non-zero $j_e^{\mu}$ and
$j_m^{\mu}$), one can obtain non-symmetrical equations (with only
one $j^{\mu}$ current) in which the resulting particles carry both
electric and magnetic charges. For a particle with explicit
magnetic charge to be observable, its ratio   $m/e$ must be
different than those of other particles.

\subsection{Global extended dual symmetry}

A more consistent dual-symmetric formalism may be defined by the
introduction of a new tensor field \bea
C^{\mu\nu}=\pa^\mu\,C^{\nu}-\pa^\nu\,C^{\mu} \label{cmunu} \eea
which is independent of $F_{\mu\nu}$ (Singleton 1995; Kato \&
Singleton 2002; Bliokh, Bekshaev \& Nori 2013). $F^{\mu\nu}$ and
$C^{\mu\nu}$ then transform as (see Eq. (\ref{max34})): \bea
F^{\,\mu\nu}&\to&\cos\theta\, F^{\mu\nu}+ \sin\theta\, C^{\mu\nu}\nn\\
C^{\,\mu\nu}&\to&-\sin\theta\, F^{\mu\nu} +\cos\theta\,
C^{\mu\nu}\,. \label{max36} \eea
 Transformations (\ref{max36}) reduce to
(\ref{max34}) when the {\it dual cons\-traint} is imposed \bea
C^{\mu\nu} \to  ^*\!\!F^{\mu\nu} \, . \eea  The  dual-symmetric
formalism acquires a particularly abbreviated form if we introduce
the complex Riemann-Silberstein four-potential $Z^{\prime \mu}$
and the field tensor $G^{\mu \nu}$
\begin{equation}
Z^{\prime \mu} = A^{\mu}+i\,C^{\mu} \nn \, ; \,\, G^{\mu\nu}=
F^{\mu\nu}+i\,C^{\mu\nu}
\end{equation}
With these definitions, instead of the Lagrangian density
(\ref{FF}) we may define
\begin{equation}
{\cal L}_0 = - \frac{1}{8}\, G^{\mu\nu} \,  G^*_{\mu\nu} \,;
\label{GG}
\end{equation}
with the duality constraint $^*G^{\mu\nu} = -i G^{\mu\nu}$, which
reduces Maxwell's equations now to a single equation
\begin{equation}
\partial_{\nu} G^{\mu\nu} = 0. \end{equation} The dual transformation
becomes now a simple $U(1)$ gauge transformation
\begin{equation}
Z^{\prime \mu} \to e^{-i \theta} Z^{\prime \mu} \, ; \, \,
G^{\mu\nu} \to e^{-i\theta} G^{\mu\nu} \, .
\end{equation}
The Lagrangian density (\ref{GG}) is invariant under this
transformation. From Noether theorem the conserved current
$J^{\mu}$ may be obtained
\begin{equation}
J^{\mu} = \frac{1}{2} Im(G^{\mu\nu}Z^{\prime *}_{\nu}) \, ; \, \,
\partial_{\mu}J^{\mu} = 0 \, ;
\end{equation}
in this expression the second equation represents helicity
conservation in Maxwell's equations.

\subsection{Dark matter}

The extension of the previous treatment to include DM is straight
forward. To this end, we introduce a complex four-vector current
\begin{equation}
J^{\mu} = j^{\mu} + i j_{\chi}^{\mu}
\end{equation}
where $j^{\mu}$ represents the Standard Model current densities,
associated to quarks and leptons, and $j_{\chi}^{\mu}$ describes
current densities of DM fermions only accessible by the extra
gauge field $C^{\mu}$. Extended dual symmetry transformations
which include DM may be synthesized as
\begin{equation}
Z^{\prime\mu} \! \to \! e^{-i \theta} Z^{\prime\mu} \, ; \, \, G^{\mu\nu} \!
\to \! e^{-i\theta} G^{\mu\nu} \, ; J^{\mu} \! \to \! e^{-i
\theta} J^{\mu} . \label{theta}
\end{equation}
The corresponding Lagrangian density may be written as
 \bea
\call&=& -\frac{1}{8}\,G^{\mu\nu}\,G^*_{\mu\nu}\,+
\frac{1}{2}\left\lbrack \,J^{\mu}\,Z_{\mu}^{\prime\ast}
+J_{\mu}^\ast\,Z^{\prime\mu} \right\rbrack\,, \label{max39} \eea where
the first term in  (\ref{max39}) is the free field Lagrangian
$\call_0$ and the second and third terms correspond to the
interaction Lagrangian density $\call_{\rm int}$.

\section{Dual Dark Matter Theory}

In the following we study the consequences of  promoting the
extended global extended version of dual symmetry (DuSy) to a
local gauge symmetry, which always come together with some boson
gauge field.
 This implies to make the substi\-tution $ \theta\to
g\,\theta(x) $ in Eq. (\ref{theta}), where $g$ represents a
coupling constant.
 Infinitesimal variations of the $Z^{\prime\mu}$ field and its partial derivatives result in
\bea \delta Z^{\prime \mu} &=& -i g \,\theta \,Z^{\prime \mu} \, ;
\, \, \delta (\pa_\nu Z^{\prime \mu})= -i g \,\pa_\nu(\theta
\,Z^{\prime \mu}) \, ;
\nn\\
\delta Z^{\prime \mu \ast} & =& i g \,\theta\, Z^{\prime \mu \ast}
\, ; \, \,  \delta (\pa_\nu Z^{\prime \mu \ast})=i g
\,\pa_\nu(\theta \,Z^{\prime \mu \ast})\,. 
\label{ldu1} 
\eea 
From these calculations it is straightforward to show that
\bea 
\delta
\, G_{\mu\nu}&=&-i g \,\theta \,G_{\mu\nu}-i g \, \lbrack
(\pa_\mu\theta)Z^\prime_\nu-(\pa_\nu\theta)Z^\prime_\mu \rbrack
\nn\\
 \delta \, G_{\mu\nu}^\ast&=& i g \,\theta \,G_{\mu\nu}^\ast + i
g \, \lbrack
(\pa_\mu\theta)Z^{\prime\ast}_\nu-(\pa_\nu\theta)Z^{\prime\ast}_\mu
\rbrack\,. 
\label{ldu2} 
\eea

To examine whether the local DuSy  is a symmetry of the gauge
field Lagrangian, the variation of $\call_0$ in (\ref{max39}) must
result in zero. This is easily calculated 
\bea 
\delta\call_0
&=&-\frac{1}{8}\, \left( \delta \, G_{\mu\nu}\, \,G^{\mu\nu\ast}
+\, G_{\mu\nu}\,\,\delta \, G^{\mu\nu\ast} \right)
\nn\\
&=&
\frac{ig}{4}
\,(\pa^{\mu}\theta)\,\left[Z^{\prime \nu }\,G^\ast_{\mu\nu} -
Z^{\prime \nu\ast}\,G_{\mu\nu} \right]\,, 
\label{ldu3} 
\eea
revealing that $\delta\call_0\neq 0$.

To assure that this symmetry is preserved, we define a
counter-term $\call_1$ and a new complex field $W^\mu$, such that
$\delta(\call_0+\call_1)=0$, resulting in 
\bea \call_1&=&
\frac{g}{4}
\,\left[
 W^\mu\,Z^{\prime \nu}\,G_{\mu\nu}^{\ast}
+ W^{\mu \ast}\,Z^{\prime\ast \nu}\,G_{\mu\nu} \right]\,.
\label{ldu4} 
\eea 
The complex field $W^\mu$ is subject to the
following transformation properties under local DuSy 
\bea 
W^\mu \to  W^\mu-i\,\pa^\mu\theta 
\, ; \,\,  W^{\mu \ast} \to W^{\mu\ast}+i\,\pa^\mu\theta\,. 
\label{ldu5} 
\eea
The variation $\delta\call_1$ results in 
\bea \delta\call_1&=&
-\frac{ig}{4}\,
(\pa^{\mu}\theta) \,\left[
 Z^{\prime\nu}\,G_{\mu\nu}^\ast-Z^{\prime \nu \ast}\,G_{\mu\nu} \right]
\nn\\
&& +
\frac{ig^2}{4}
\,(\pa_{\mu}\theta)\left[
W^\mu\,Z^{\prime\nu}\,Z^{\prime\ast}_{\nu}
-
W^{\mu\ast}\,Z^{\prime\nu\ast}\,Z^\prime_{\nu}\right]
\nn\\
&& +
\frac{ig^2}{4}
\,(\pa_{\mu}\theta)\left[
Z^{\prime\mu\ast}\,W^{\nu\ast}\,Z^\prime_{\nu}
-
Z^{\prime\mu}\,W^{\nu}\,Z^{\prime\ast}_{\nu}
\right]\,.
\label{ldu6} 
\eea 
The first term in (\ref{ldu6})
cancels $\delta\call_0$, but two extra terms proportional to $g^2$
appear. Now in order to  cancel the second term in (\ref{ldu6}), a
new a counter-term $\call_2$ must be introduced: 
\bea 
\call_2&=& -\frac{g^2}{4} \,W_\mu W^{\mu\ast} \,Z^\prime_\nu Z^{\prime\nu\ast}\,.
\label{ldu7a} 
\eea 
The variation $\delta\call_2$ produces
{\setlength{\mathindent}{0pt}
\bea 
\delta\call_2&=& 
-\frac{g^2}{4}
\left[
(\delta\,W_\mu) W^{\mu\ast}
\,Z^\prime_\nu Z^{\prime\nu\ast}\, 
-\,W_\mu
(\delta\,W^{\mu\ast}) Z^\prime_\nu Z^{\prime\nu\ast}\,
\right.
\nn\\
&&
\left.
-\,W_\mu W^{\mu\ast} \delta\left(\,Z^\prime_\nu Z^{\prime\nu\ast}\right)
\right].
\label{ldu7b}
\eea
}
It is easy to prove that \bea \delta\left(\,Z^\prime_\nu
Z^{\prime\nu\ast}\,\right)&=&(\delta\,Z^\prime_\nu)
\,Z^{\prime\nu\ast}\, +Z^\prime_\nu\,
(\delta\,Z^{\prime\nu\ast})\,
\nn\\
&=& -i\,g\,\theta\,Z^\prime_\nu
Z^{\prime\nu\ast}+i\,g\,\theta\,Z^\prime_\nu Z^{\prime\nu\ast} \nn
= 0\,, \label{ldu7c} 
\eea 
which shows that the last term in Eq.
(\ref{ldu7b}) is zero. Using Eq. (\ref{ldu5}), one obtains for the
remaining terms of Eq. (\ref{ldu7b}) 
\bea 
\delta\call_2&=&
 -\frac{ig^2}{4}\,
(\pa_{\mu}\theta)
\left[
W^\mu\,-W^{\mu\ast}
\right]
Z^{\prime\nu}\,Z^{\prime\ast}_{\nu} \,,
\label{ldu7d} 
\eea
which cancels the second term in Eq. (\ref{ldu6}). 
The remaining  third  term in Eq. (\ref{ldu6}) is
cancelled with new a counter-term $\call_3$: 
\bea \call_3&=&
\frac{g^2}{8} \left[ \frac{}{} W^\nu W_\mu^\ast Z^{\prime\ast}_\nu
Z^{\prime\mu}
+W^{\nu\ast} W_\mu  Z^\prime_\nu  Z^{\prime\mu\ast} \right]\,.
\label{ldu7e} 
\eea 
The variation of the first term in Eq.
(\ref{ldu7e}) results in 
\bea
\delta( W^\nu W_\mu^\ast Z^{\prime\ast}_\nu Z^{\prime\mu})&=&
i\,\pa_\mu\,\theta\, W^\nu Z^{\prime\ast}_\nu Z^{\prime\mu}
\nn\\&& - i\,\pa_\mu\,\theta\,W^{\nu\,\ast} Z^{\prime}_\nu
Z^{\prime\mu\ast}
\nn\\
&& + W^\nu W_\mu^\ast\, \delta( Z^{\prime\ast}_\nu
Z^{\prime\mu})\,. 
\label{ldu7f} 
\eea 
From (\ref{ldu7c}), the last term in Eq. (\ref{ldu7f}) is zero, resulting in
\bea 
\delta( W^\nu W_\mu^\ast Z^{\prime\ast}_\nu Z^{\prime\mu})&=&
i\,\pa_\mu\,\theta\, W^\nu Z^{\prime\ast}_\nu Z^{\prime\mu}
\nn\\&& -i\,\pa_\mu\,\theta\, W^{\nu\,\ast} Z^{\prime}_\nu
Z^{\prime\mu\ast}\,. 
\label{ldu7g} 
\eea 
The variation of the second term in Eq.  (\ref{ldu7e}) is obtained from Eq.
(\ref{ldu7g}) by exchanging $\mu\leftrightarrow\nu$ 
\bea 
\delta(
W^\mu W_\nu^\ast Z^{\prime\ast}_\mu Z^{\prime\nu})&=&
i\,\pa_\nu\,\theta\, W^\mu Z^{\prime\ast}_\mu Z^{\prime\nu}
\nn\\&& -i\,\pa_\nu\,\theta\, W^{\mu\,\ast} Z^{\prime}_\mu
Z^{\prime\nu\ast}\,. 
\label{ldu7h} 
\eea 
Combining (\ref{ldu7e}), (\ref{ldu7g}) and (\ref{ldu7h}) allows one to calculate
$\delta\call_3$ 
\bea 
\delta\,\call_3=-\frac{ig^2}{4}
\,(\pa_{\mu}\theta)\left[
Z^{\prime\mu\ast}\,W^{\nu\ast}\,Z^\prime_{\nu} -
Z^{\prime\mu}\,W^{\nu}\,Z^{\prime\ast}_{\nu} \right]\,
\label{ldu7i} 
\eea
which cancels the third term in (\ref{ldu6}).
The total invariant DuSy Lagrangian corresponds to the sum $
\call_{\rm DuSy}=\call_0+\call_1+\call_2+\call_3\,, 
$ which can be written in a very compact form
\bea
\call_{\rm DuSy}&=&-\frac{1}{8}|G_{\mu\nu}-g K_{\mu\nu}|^2
+m_Z^2\,|Z^\prime_\mu|^2
\nn\\
&&
-\frac{1}{2} F_{ W \mu\nu} F_W^{\mu\nu\ast}+|m_W\,W_\mu-\pa_\mu\sigma|^2
\label{ldu8}
\eea
where 
\bea 
K_{\mu\nu}=Z^\prime_\mu W_\nu-Z^\prime_\nu W_\mu \,  ; \, \,
F_{ W \mu\nu}=\pa_\mu W_\nu-\pa_\nu W_\mu\,. \label{ldu8b} 
\eea 
\hspace{-.2cm}
The transformation of  $F_{ W \mu\nu}$ under DuSy transformation (\ref{ldu5}) is
invariant, as can be seen
\bea \!\!\!\!\!\!\!\!F_{ W \mu\nu} \!\to \!  \pa_\mu ( W_\nu-i
\pa_\nu\theta)\!-\!\pa_\nu ( W_\mu-i \pa_\mu\theta) = F_{ W
\mu\nu} \, , \label{ldu8c} \eea
which proves the invariance of $-\frac{1}{2} F_{ W \mu\nu}
F_W^{\mu\nu\ast}$ term in (\ref{ldu8}). The two mass terms where
included in (\ref{ldu8}) for the  $Z^\prime_\mu$ and $W_\mu$
bosons, which at a first glance seem to break the DuSy symmetry.
The $m_Z$ mass term depends on $|Z^\prime_\mu|^2$  which is
trivially invariant under the dual transformation.   The
invariance of the $m_W$ mass term is not so trivial. It requires
the introduction of a complex $\sigma$  Stueckelberg field in
(\ref{ldu8}).  Stueckelberg's wonderful trick relies in the
introduction of an extra
 scalar field $\sigma$  (in our case a complex scalar), in addition to the four components $W_\mu$.
This  results in  a total of five fields, to describe covariantly
the three polarizations of a massive vector field. The
Stueckelberg mechanism, not only  manifests  Lorentz covariance,
but also, gauge invariance. In this sense the Stueckelberg field
restores the gauge symmetry which had been broken by the mass term
(Ruegg \&,Ruiz-Altaba 2004), (K\"ors \& Nath 2004, 2005),
(Feldman,  K\"ors \& Nath 2007), (Cheung \&  Yuan 2007), (Zhang,
Wang, \& Wang 2008). Under dual transformation $W_\mu$ transforms
according to (\ref{ldu5}) and $\sigma$ as \bea
\sigma\to\sigma-i\,m_W\,\theta\,. \label{ldu9} \eea Under  the
transformations (\ref{ldu5}) and (\ref{ldu9}), the mass term  of
the $W_\mu$ field transforms as \bea m_W W_\mu-\pa_\mu\sigma &\to&
m_W (W_\mu-i\,\pa_\mu\theta)
-\pa_\mu(\sigma-i\,m_W\,\theta\,) \nn\\&=& m_W W_\mu-\pa_\mu\sigma
\label{ldu9b} \eea proving its invariance under DuSy
transformation. The $W_\mu$ and $\sigma$ fields decouple if one
introduces an extra gauge fixing term to the Lagrangian
(\ref{ldu8}) \bea \call_{\rm
gf}=-\frac{1}{\xi}\,|\pa_\mu\,W^\mu+\,\xi\,m_W\sigma|^2\,.
\label{ldu10} \eea Then, the $\sigma$ field decouples as we can
see by the following development {\setlength{\mathindent}{0pt}
\bea |\!\!&m_W\!\!& W_\mu \! -\pa_\mu\sigma|^2 + \call_{\rm gf}=
m_W^2\,|W_\mu|^2 +\frac{1}{\xi}|\pa_\mu\,W^\mu|^2
\nn\\
&-&
 m_W(\, W^\mu\,\pa_\mu\sigma^\ast+W^{\mu\ast}\,\pa_\mu\sigma\,)
\nn\\
&+&
 m_W(\, W^\mu\,\pa_\mu\sigma^\ast+W^{\mu\ast}\,\pa_\mu\sigma\,)
\\
&+&
|\pa_\mu\,\sigma|^2-\xi\,m_W^2\,|\sigma|^2
\nn\\
&=&m_W^2\,|W_\mu|^2 +\frac{1}{\xi}|\pa_\mu\,W^\mu|^2 +
|\pa_\mu\,\sigma|^2-\xi\,m_W^2\,|\sigma|^2 . \nn \label{ldu10b}
\eea } \hspace{-.2cm} The mixing term $\sigma \pa_\mu W^{\mu\ast}$
that appeared in $\call_{\rm gf}$ cancels the corresponding mass
term. This leads to the decoupling of the auxiliary complex scalar
$\sigma$ and the vector field $W_\mu$. A mass proportional to a
random parameter $\sqrt{\xi}$
 is then given to the
unphysical $\sigma$ field which has no influence on the vector
field $W_\mu$.
Finally the complete Lagrangian  in {\it Dual DM}  theory is given
by
 \bea
\!\!\!\! \call_{\rm DDM}&=& \call_{\rm DuSy}\,+ \call_{\rm gf}+
\frac{1}{2}\left\lbrack \,J^{\mu}\,Z_{\mu}^{\prime\,\ast}
+J_{\mu}^\ast\,Z^{\prime\mu} \right\rbrack \, . \label{ldu11} \eea

As we have seen, promoting dual symmetry to a local symmetry for
the the complex  four-potential $Z^\prime_\mu$   implies in the
inclusion of an extra complex vector field $W_\mu$ with a complex
gauge transformation. Our conjecture  is the following:  an
interesting and possible connection of this dark sector with the
SM can be established if one assumes that $W_\mu$ represents the
electroweak charged gauge bosons $W^{\pm}_\mu$. This assumption is
not in contradiction with the known fact that $W^{\pm}_\mu$ is
part of the $SU(2)$ electroweak symmetry. As a novelty, we assume
that $W^{\pm}_\mu$ has an additional dual symmetry associated to
$Z^\prime_\mu$.

\subsection{Perspectives in a new CP violation scenario}

If the extended dual symmetry is a fundamental symmetry of
physics, then our tentative conjecture  to interpret $W_\mu$ as
the weak interaction charged vector field $W^{\pm}_\mu$, leads us
to speculate about a possible extra CP violation\footnote{As is
well  known, the CP trans\-for\-mation combines charge
conjugation, re\-pre\-sen\-ted by the operator C with the parity
transformation symbolized by P. Under charge transformation C,
particles and antiparticles are interchanged, by conjugating all
internal quantum numbers, as for instance $Q \to - Q$ in case of
electromagnetic charge. Under parity operation P, the handedness
of space is reversed, $\vec{x} \to - \vec{x}$. In this case, a
left-handed electron for instance $e^{\tiny -}_L$ is transformed
under CP into a right-handed positron, $e^{\tiny +}_R$. CP could
be violated also by the strong interaction. However, the smallness
of  the non-perturbative parameter associated to the strength CP
violation in QCD, $\theta_{QCD}$, constitutes a theoretical
puzzle, known as {\it the strong CP problem}.} in electroweak
interaction. In the SM, quark mixing is the  source of CP
violation. The charged current interactions for quarks, i.e.,  the
$W^{\pm}$ interactions, are given by
 \bea
-\call_{W^{\pm}}=\frac{g'}{\sqrt{2}}\,\,\overline{\calu}_{Li}\,\gamma^\mu
(V_{\rm CKM})_{ij} \,\cald_{Lj}\,W^{\pm}_\mu +{\rm h.c.} \, ,
\label{lw} \eea where the The Cabibbo-Kobayashi-Maskawa mixing
matrix is defined as
 \bea V_{\rm CKM} =
\left(%
\begin{array}{ccc}
V_{ud}   & V_{us} & V_{ub} \\
 V_{cd}   & V_{cs} & V_{cb} \\
V_{td}   & V_{ts} & V_{tb}  \\
\end{array}%
\right) \,.
\label{ckm}
\eea
While a general $3 \times 3$ unitary matrix depends on three real
angles and six phases, the freedom to redefine the phases of the
quark mass eigenstates can be used to remove five of the phases,
leaving a single physical phase, the Kobayashi-Maskawa phase, that
is responsible for all CP violation in meson decays in the
Standard Model.
As a
consequence  of this fact  $V_{\rm CKM}$ is not diagonal and the
$W^{\pm}_\mu$ gauge bosons couple to quark (mass eigenstates) of
different generations. One can say that within the Standard Model,
this is the only source of flavor changing interactions.
 
\begin{figure}
\begin{center}
\includegraphics[width=30mm,height=30mm]{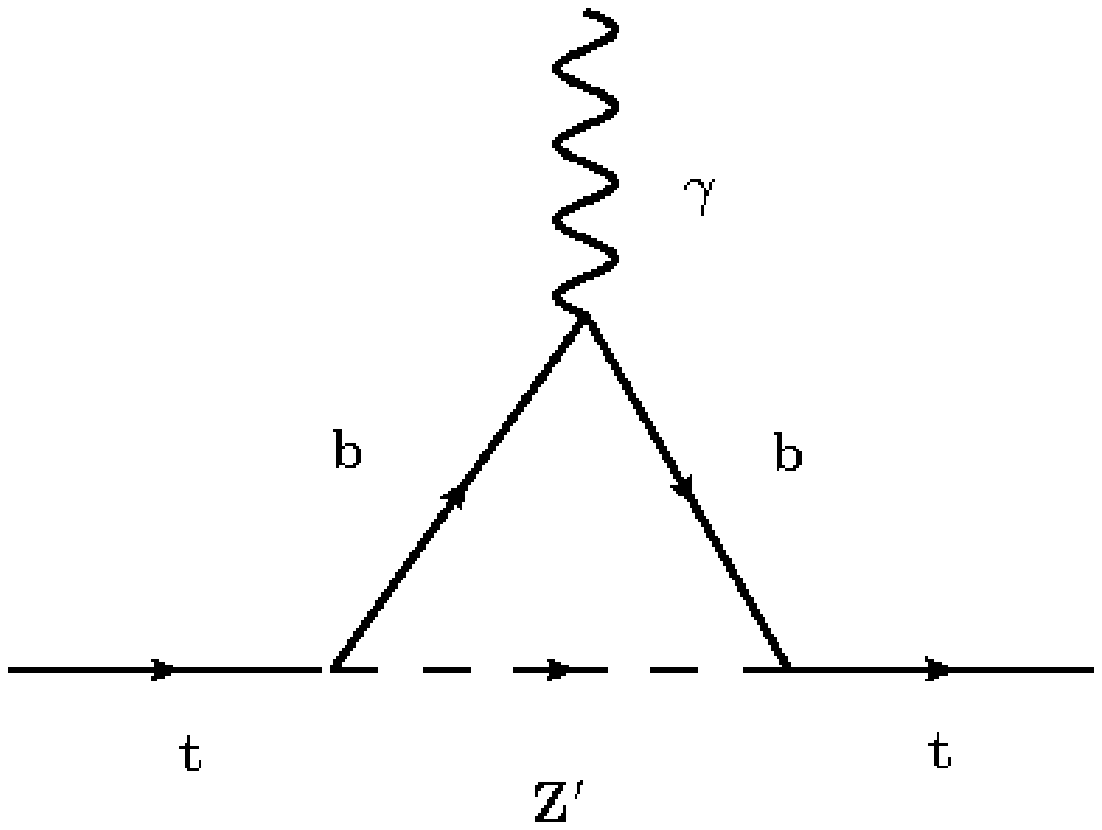}
\includegraphics[width=30mm,height=30mm]{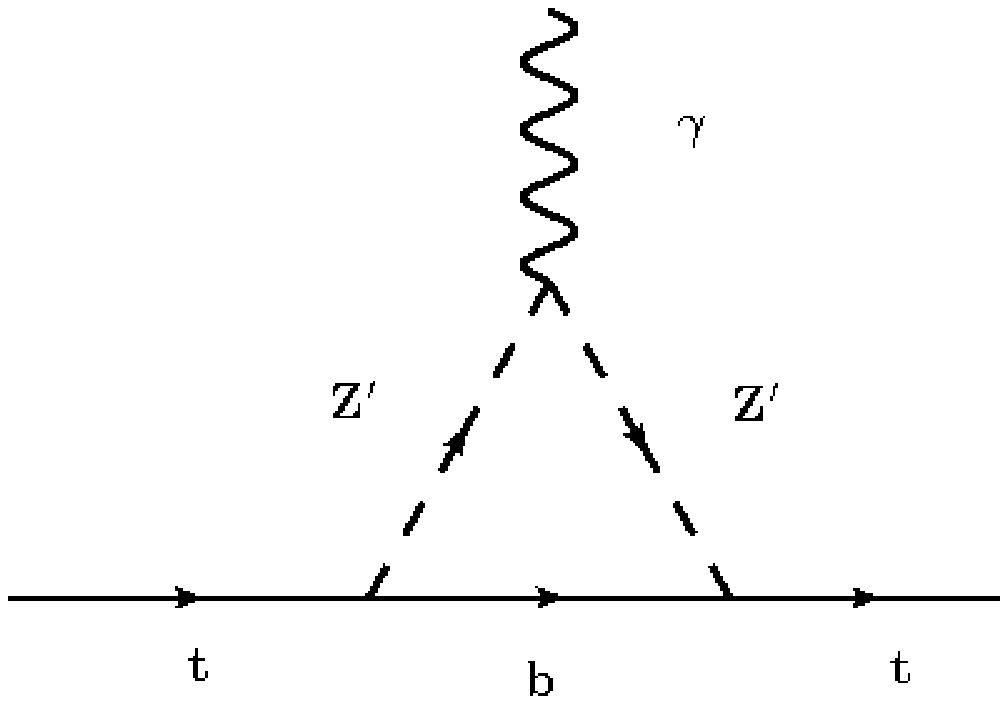}
\vspace{-0.5cm} \caption{Feynman one loop correction to
$\bar{t}t\gamma$} \label{fig1}
\end{center}
\end{figure}

Consider in the following  particles  $A$ and $B$ and their
antiparticles $\bar{A}$ and $\bar{B}$ for which to the general
symmetry transformation process $A\to B$ corresponds an
antiparticle transformation symmetry process $\bar{A}\to\bar{B}$.
We denote the amplitudes of these transformation processes as
$\calm$ and $\bar{\calm}$ respectively. If CP violation occurs in
these processes,  these terms must correspond to the same complex
number. We can separate the magnitudes and phases associated to
these processes by writing $\calm=|\calm|e^{i\theta}$. If a phase
term is introduced from (e.g.) the CKM matrix, call  it $e^{i\phi}
$, then we have the
two amplitudes written as: 
$\calm = |\calm|e^{i\theta}e^{i\phi}; \,
\bar{\calm}=|\bar{\calm}|e^{i\theta}e^{-i\phi}\,.$
 Now, consider that
there are two different decay  channels  for $A\to B$:
$\calm=|\calm_1|e^{i\theta_1}e^{i\phi_1}+|\calm_2|e^{i\theta_2}e^{i\phi_2};
\,
\bar{\calm}=|\calm_1|e^{i\theta_1}e^{-i\phi_1}+|\calm_2|e^{i\theta_2}e^{-i\phi_2}.$
Combining these expressions, in a CP violation scenario we get
\beann 
|\calm|^2\!-\!|\bar{\calm}|^2=- 4 |\calm_1||\calm_2|
\sin(\theta_1-\theta_2)\sin(\phi_1-\phi_2),
\eeann 
where a complex phase gives rise to processes that
proceed at different rates for particles and antiparticles. As a
text of this picture, the presence of the complex $Z^\prime_\mu$
could lead to an extra phase dependence that in future
calculations could be evaluated. Another test could be, for
instance,  the one related to a very similar SM interaction, with
top and botton quarks, described by the corresponding $Wtb$
vertex. In this case, one can write the equivalent interaction of
the $Z^\prime_\mu$ field with top and bottom quarks as (Najafabadi
2010): 
\bea 
-\call=\frac{g}{\sqrt{2}}\,\,\overline{t}\,\gamma^\mu
\, \left(a_L\,P_L+a_R\,P_R \right)
 \,b\,Z^\prime_\mu
+{\rm h.c.}
\label{lzp} 
\eea 
where $P_L$ ($P_R$ ) are the left-handed
(right-handed) projection operators;  $a_L$ and $a_R$ are the
complex coefficients, where  CP violating effects will appear.
This interaction of the complex $Z^\prime_\mu$ with top and bottom
quarks consists of $V-A$ and $V+A$ structures and could produce an
electric dipole moment  for the top quark at one loop level,
considering the Lagrangian defined in (\ref{lzp}).  The relevant
Feynman diagrams are shown in figures (\ref{fig1}). This could be
one of the many scenarios where this complex $Z^\prime_\mu$ theory
could be tested.

%

\section{Conclusions}

In this paper   we have studied the consequences of extending dual
symmetry to include a generic $C_\mu$ vector field as a dual
partner of the photon $A_\mu$ and represented these fields by a
complex Riemann-Silberstein-like four-potential $Z^\prime_\mu$.
The promotion of dual symmetry to a local symmetry for the the
complex  four-potential $Z^\prime_\mu$   implies in the inclusion
of an extra complex vector field $W_\mu$ with a complex gauge
transformation. A  Dual DM Lagrangian $\call_{\rm DDM}$ is
obtained from the general DuSy invariant Lagrangian $\call_{\rm
DuSy}$. Our tentative conjecture is to interpret $W_\mu$ as the
actual weak interaction charged $W^{\pm}_\mu$, lead us to
speculate about a possible extra CP violation scenarios for future
calculations.

\vspace{.5cm}
\acknowledgements

 This research was supported by
Conselho Nacional de Desenvolvimento Cient\'{\i}fico e
Tecnol\'ogico (CNPq).


\end{document}